\documentstyle[aps,preprint,prl,12pt,amsfonts,epsf,rotate]{revtex}
\epsfverbosetrue

\newif\iffirstfig \global\firsttabtrue

\newcommand{\vv}{\vspace{1cm}}
\newcommand{\beq}{\begin{equation}}
\newcommand{\eeq}{\end{equation}}
\newcommand{\beqn}{\begin{eqnarray}}
\newcommand{\eeqn}{\end{eqnarray}}
\newcommand{\beqa}{\begin{eqnarray}}
\newcommand{\eeqa}{\end{eqnarray}}
\newcommand{\bi}{\bibitem}
\newcommand{\n}{\newline}
\newcommand{\nn}{\nonumber\\}
\newcommand{\bc}{\begin{center}}
\newcommand{\ec}{\end{center}}
\newcommand{\ov}{\overline}
\newcommand{\la}{\langle}
\newcommand{\lla}{\left\langle}
\newcommand{\rra}{\right\rangle}
\newcommand{\ra}{\rangle}
\newcommand{\cl}{{\cal C}}
\newcommand{\gl}{{\cal G}}

\begin{document}
\pagestyle{empty}
\title{
On the out of equilibrium order parameter in long-range spin-glasses.
}
\author{
A. Baldassarri,
L. F. Cugliandolo
\footnote{
Address after October 1st: Service de Physique de l'Etat Condens\'e,
CEA, Saclay, France.
},
J. Kurchan
\footnote{
Address after October 1st: LPTHE, Ecole Normale Sup\'erieure, Paris, France.
},
G. Parisi
}
\address{
Dipartimento di Fisica,
Universit\`a di Roma I, {\it La Sapienza}, Roma, Italy,
\\
INFN - Sezione di Roma I
}

\maketitle

\begin{abstract}
We show that the  dynamical order parameters can be reexpressed
in terms of the distribution of the staggered auto-correlation
and response functions.
We calculate these distributions for the out of equilibrium dynamics of the
Sherrington-Kirpatrick model at long times.
The results  suggest that the landscape this model visits
at different long times
in an out of equilibrium relaxation process  is, in a sense, self-similar.
Furthermore, there is a
similarity between the landscape seen out of equilibrium at
long times and the equilibrium landscape.

The calculation is greatly simplified by making use of the superspace
notation in the  dynamical approach. This notation also highlights the
rather mysterious formal connection between the dynamical and replica
approaches.

We also perform numerical simulations which show good agreement with
the analytical results for the out of equilibrium dynamics.

\end{abstract}
\vspace{1cm}

\newpage

\pagestyle{plain}

\section{Introduction}
\renewcommand{\theequation}{\thesection.\arabic{equation}}
\setcounter{equation}{0}
\label{sec1}

The partition function of mean-field  spin-glasses
is dominated by many states. The geometrical organization of these states,
their relative weights in the Gibbs-Boltzmann measure, and the
distribution of their mutual distances
have been known  for some time \cite{Pa,Mepavi}.
Of particular importance is the functional order parameter
$P(q)$ giving the probability distribution of states with
mutual overlap $q$.

The Gibbs-Boltzmann measure can be studied analytically using a dynamical
approach \cite{Sozi}. For instance, the Langevin Dynamics
\beq
\Gamma_0^{-1} \, \partial_t \sigma_i(t)
=
- \beta \frac{\delta H}{\delta \sigma_i(t)}
+ \xi_i(t)
\;
\label{lang}
\eeq
($\Gamma_0$ determines the time scale and
$\xi_i(t)$ is a Gaussian white noise with zero mean and variance $2$),
with the following order of large times and large $N$ limits
\beq
  \lim_{N \rightarrow \infty} \;\;\; \lim_{t \rightarrow \infty}
\; ,
\label{lim2}
\eeq
guarantees ergodicity
and leads the system to equilibrium.
The equilibrium thermodynamical values of any operator $O$ are then
obtained as averages over the noise
$\la O \ra_{eq}
=
\lim_{N\rightarrow\infty} \lim_{t\rightarrow\infty}  \la O(t) \ra $.

\vv

A different situation, closer to the experimental settings, is to consider
the relaxation of an infinite system at long but finite times.
The time is measured from the initial time - the quenching
time in experiments - which we take as zero. The order of limits is then
\beq
\lim_{t \rightarrow \infty} \;\;\; \lim_{N \rightarrow \infty}
\label{lim1}
\; .
\eeq

An analytical solution for mean-field spin-glasses
in this regime  has been recently developed
\cite{Cuku1,Frme,Cuku2,Frme2}.
It was there argued that in regime
(\ref{lim1}) mean-field spin-glass systems {\em below} the critical temperature
never achieve equilibrium, not even within a restricted sector of
phase-space. This is in agreement with experimental spin-glasses, for which
the estimate is that aging effects take a few years to die away \cite{Boviha}.

\vv

A relevant order parameter for the long-time asymptotics of
the relaxation is
the  dynamical $P_d(q)$ defined as follows \cite{Cuku1}:
We add  time-independent source terms   $h_{i_1 \dots i_{r}}$ to the energy
\beq
H_h = H +
\frac{1}{N^{r-1}}
\sum_{i_1 , \dots , i_r}^N
h_{i_1  \dots i_r}
\sigma_{i_1} \dots \sigma_{i_{r}}
\; ,
\nonumber
\eeq
and then consider
the generating functions of the generalized susceptibilities
\beqa
Lim \;\;
\left[ 1
-
\frac{r}{N^r}
\sum_{i_1 < \dots < i_r}
\left.
{
\frac{\partial
\overline{
<
s_{i_1}(t) \dots s_{i_r}(t)
>
}
}
{
\partial
h_{i_1 \dots i_r}
}
}\right|_{h=0} \right]
\hspace{2cm}
\nn
\hspace{1cm}
=
\int_0^1 dq' \, \frac{d X_{(\,)}(q')}{ d q'}
\, {q'}^r=\int_0^1 dq' \, P_{(\,)} (q') \, {q'}^r
\; .
\label{aa}
\eeqa
If the sysmbol $Lim$ stands for (\ref{lim2}),
it defines the usual Parisi order
parameters $x(q)$ and $P(q)$ \cite{Pa}.
If it stands for (\ref{lim1}) then Eq. (\ref{aa})
defines the dynamical parameters $X_d(q)$ and $P_d(q)$
\cite{Cuku1}.

The   dynamical order parameters $X_d(q)$ and $P_d(q)$
- unlike their static counterparts - have not as yet been
given a probabilistic interpretation. The main purpose of
this paper is to show that $P_d(q)$ can be recast into a form that:

{\it i.} Makes its physical meaning more explicit.

{\it ii.} Shows for the Sherrington-Kirkpatrick (SK) model that their
is a self-similarity in the landscape.
Although at all long but finite times (limit (\ref{lim1})) the system is
exploring regions of phase-space which it will eventually leave,
never to return, some
geometrical properties of these regions coincide with those of the
equilibrium states.

{\it iii.} Is amenable to numerical simulations.

\vv

The Hamiltonian of the Sherrington-Kirpatrick  model
is given by
\begin{equation}
H(\sigma)
=
\frac{1}{N}
\sum_{ij}
J_{ij} \sigma_{i} \sigma_{j} +a \sum_i   \; (\sigma_i^2-1)^2
\; .
\label{1.1}
\end{equation}
The interactions $J_{ij}$ are quenched random variables
Gaussianly distributed with  zero mean and
variance $1/\sqrt N$.
The last is a spin weight term and the hard-spin limit ($\pm 1$)
is recovered taking $a \rightarrow \infty$.

The out of equilibrium dynamics of the SK model has been studied in
Ref. \cite{Cuku2}.
A surprising outcome, obtained under a set of hypothesis
described there in detail, is that the dynamical order parameters
$P_d(q)$ and $X_d(q)$
coincide with the static order parameters $P(q)$ and $X(q)$
- even if the physical situations they describe are very different.
The coincidence of dynamical and static order parameters does
not hold for every model, for instance the $p$-spin
spherical model behaves in a different way \cite{Cuku1}.

We shall use in this paper the results of Ref. \cite{Cuku2}
to compute analytically the staggered auto-correlation and response
functions \cite{Daso,Mepa}
in the limit of any two large times (limit (\ref{lim1}))
for the SK model.
All the information about the asymptotic large times solution is encoded
in the order parameter $X_d(q)$ and in the `triangle relation' $f$
relating the correlation functions at any three long times (see Ref.
\cite{Cuku2}).
We shall
then compare analytical and numerical results for the staggered distributions.
Their good agreement
gives numerical support for the predicted equality of
dynamical and static order parameters in this case.

\vv

In order to obtain the staggered distributions from $X_d(q)$ (or $P_d(q)$)
and $f$
we shall heavily use the {\it formal} relation between the static
replica approach and
the dynamical approach, which becomes transparent  when the latter
is formulated in terms of superspace variables \cite{Ku,Cufrkume}.
(Although the   underlying supersymmetry in this dynamics is partially
broken by the `boundary' - initial - conditions, it  still has
useful consequences.)

This proceeds in two steps: we firstly identify the dynamical
- superspace - counterparts of the static - replica space - variables.
We obtain, roughly speaking, the same formul\ae $\,$ with superspace
integrals  - including a time-integral - substituting sums over replicas.

Secondly, we look for the solution for the dynamical order parameter.
Again, this solution has many points in common with the solution for
the statics although they are not  equivalent for every model and describe
entirely different physical situations.

\vv

The paper is organised as follows.
In Section \ref{sec2} we introduce the staggered distributions.
Using the SUSY formalism, we find that they are related
to the powers of the dynamical order parameter ${\bbox Q}(1,2)$.
In Section \ref{sec3}
we compute, in general, these powers of
${\bbox Q}(1,2)$ and then specialise to the long-time asymptotics of
the SK model using the results of Ref. \cite{Cuku2}.
In Section \ref{sec4} we obtain the staggered auto-correlation function
for large times.
In Section \ref{sec5} we describe the numerical simulations and compare
them to the analytical results.
Finally, we discuss the physical picture in Section \ref{sec6}.

\vv
\section{Staggered distribution functions}
\renewcommand{\theequation}{\thesection.\arabic{equation}}
\setcounter{equation}{0}
\label{sec2}
\vv

The staggered auto-correlation function is defined as
\beq
g(\lambda;t_1,t_2)
\equiv
\la
\sigma_\lambda(t_1) \; \sigma_\lambda(t_2)
\ra
=
\sum_{ij}
\la \lambda | i \ra \;
\la \lambda | j \ra \;
\la \sigma_i(t_1) \; \sigma_j(t_2) \ra
\;.
\label{stagcorr}
\eeq
$\lambda$ denotes the eigenvalues of the
$N \times N$ random matrix $J_{ij}$ associated with the eigenvectors
$| \lambda \ra$.
$|\sigma(t) \ra$ is the time-dependent
$N$-dimensional vector of spins,
$\sigma_i(t) \equiv \la i | \sigma(t) \ra$, and
$\sigma_\lambda(t) \equiv \la \lambda | \sigma(t) \ra$
are the staggered spin states.

The staggered response function is
\beq
\hat{g}(\lambda;t_1,t_2)
\equiv
\lla \frac{\delta\sigma_\lambda(t_1)}{\delta h_\lambda(t_2)} \rra
=
\sum_{ij}
\la \lambda | i \ra \;
\la \lambda | j \ra \;
\lla \frac{\delta\sigma_i(t_1)}{\delta h_j(t_2)} \rra
\; .
\label{stagresp}
\eeq

The functions $g$ and $\hat{g}$ are in turn related to the set of
time-dependent two-point functions
\beqa
E^{(k)}(t_1,t_2)
&\equiv&
\frac{1}{N} \sum_{ij}
\ov{
\lla \sigma_i(t_1) \, \left(J^k\right)_{ij} \, \sigma_j(t_2) \rra
}
\; ,
\\
\hat{E}^{(k)}(t_1,t_2)
&\equiv&
\frac{1}{N} \sum_{ij}
\ov{
\lla  \left(J^k\right)_{ij} \,
\frac{\delta\sigma_i(t_1)}{\delta h_j(t_2)}
\rra
}
\; ,
\label{EE}
\eeqa
where $\ov{ \; \cdot \;}$
represents the mean over the quenched disorder.
In terms of $g$ and $\hat{g}$ they read
\beqa
E^{(k)}(t_1,t_2)
&\equiv&
\int d\lambda \, \rho(\lambda) \, \lambda^k \,
g(\lambda;t_1,t_2)
\; ,
\label{ek}
\\
\hat{E}^{(k)}(t_1,t_2)
&\equiv&
\int d\lambda \, \rho(\lambda) \, \lambda^k \,
\hat{g}(\lambda;t_1,t_2)
\; .
\eeqa
$\rho(\lambda)$ is the eigenvalue distribution
that, in the limit of large
$N$, corresponds to the semicircle law
$\rho(\lambda) = 1/(2\pi) \, \sqrt{ 4 - \lambda^2}$ if the variance
of the $J_{ij}$ is finite \cite{Me}. In particular, if $t_1 = t_2$ and $k=1$
Eq. (\ref{ek}) gives the time-dependent energy density.

\vv
\subsection{Supersymmetric Formalism}
\vv

Following Ref. \cite{Ku} we introduce the supersymmetric
`field' ${\bbox \phi}_i(1)$, $i=1,\dots,N$,
\beq
{\bbox \phi}_i(1)
\equiv
\sigma_i(t_1) +
\eta_i(t_1) \, \ov \theta_1 + \theta_1 \, \ov \eta_i(t_1)
+
\hat{\sigma}_i(t_1) \, \ov \theta_1 \theta_1
\eeq
with $1 \equiv (t_1,\theta_1, \ov \theta_1)$.

The dynamical expectation value of a quantity $O$ can then be written as
\beqa
\la O (t_1) \ra
&=&
\int \Pi_i D[{\bbox \phi}_i] \; O (t_1)
\exp [-S_{KIN} - S_{POT}]
\; ,
\nn
S_{KIN}
&=&
\Gamma^{-1}_0
\int
d\theta d\bar \theta dt \;
\sum_i \; \frac{\partial
{\bbox \phi}_i}{\partial \theta}\,
\left(
\frac{\partial {\bbox \phi}_i}{\partial \bar \theta} -
\theta
\frac{\partial {\bbox \phi}_i}{\partial t}
\right)
\; ,
\nn
S_{POT}
&=&
\beta \int d\theta d\bar \theta dt \; H({\bbox \phi})
\; .
\label{1.9}
\eeqa
As in the static replica approach, once the mean is taken over the couplings
one ends up with a functional of the order parameters that can be calculated
by saddle point. The dynamical order parameter is the
 `supercorrelation' function  defined as
\beq
{\bbox Q}(1,2) \equiv \frac{1}{N} \sum_i \; \la {\bbox \phi}_i(1)
{\bbox \phi}_i(2) \ra
\;
\eeq
that plays the same role as $Q_{ab}$ in the statics.
For the mean-field case that satisfies causality the saddle-point value
of ${\bbox Q}(1,2)$ can be
written as
\beq
{\bbox Q}(1,2)
=
C(t_1,t_2) +
(\ov \theta_2 - \ov \theta_1) \;
\left[
\theta_2 \,
\; G(t_1,t_2)
+ \theta_1 \, \; G(t_2,t_1)
\right]
\label{Q12}
\eeq
and it encodes the two-time
functions $C$ and $G$  that are
the stantard auto-correlation and response functions
\beqa
C(t_1,t_2)
&\equiv&
\frac{1}{N}  \sum_i \; \la \sigma_i(t_1) \; \sigma_i(t_2) \ra
\; ,
\\
G(t_1,t_2)
&\equiv&
\frac{1}{N}  \sum_i \; \lla\frac{\delta \sigma_i(t_1)}{\delta h_i(t_2)} \rra
\; ,
\eeqa
respectively. Because of causality, $G(t_1,t_2) = 0$ if $t_2 > t_1$.
In these formul\ae $\,$ we have ommited
the mean over the disorder since $C$ and $G$ are
self-averaging in the limit $N \rightarrow \infty$ {\em for finite times} as
can be easily proven by considering the evolution of two independent
copies of the system with the same couplings $J_{ij}$.

We shall need the definition of the operator powers of ${\bbox Q}$
\beq
{\bbox Q}^k(1,3) \equiv \int d2 \; {\bbox Q}^{k-1}(1,2) \, {\bbox Q}(2,3)
\; .
\label{powers}
\eeq
It is easy to see that ${\bbox Q}^k$ conserves the form (\ref{Q12})
with $C^{(k)}$ and $G^{(k)}$  given inductively by
\beqa
C^{(k)}(t_1,t_3)
&=&
\int dt_2 \;
\left[
C^{(k-1)}(t_1,t_2) \, G(t_3,t_2)
+
G^{(k-1)}(t_1,t_2) \, C(t_2,t_3)
\right]
\; ,
\label{ceq}
\\
G^{(k)}(t_1,t_3)
&=&
\int dt_2 \;
G^{(k-1)}(t_1,t_2) \, G(t_2,t_3)
\; ,
\label{geq}
\eeqa
$t_1 > t_3$.
{}From now on supra-indices within parenthesis denote entries
in the function ${\bbox Q}^k$ while supra-indices without parenthesis
denote ordinary powers.

\vv
\subsection{Staggered auto-correlation and response functions}
\vv

We now start the computation of the staggered auto-correlations and responses.
With the superspace notation most of the manipulations of Ref. \cite{Mepa}
carry through without change, just substituting replica indices by
superspace variables. Defining
\beq
{\bbox E}^{(k)}(1,2) =  E^{(k)}(t_1,t_2) +
(\ov \theta_2 - \ov \theta_1) \;
\left[
\theta_2 \,
 \hat{E}^{(k)}(t_1,t_2)
+ \theta_1 \,
 \hat{E}^{(k)}(t_2,t_1)
\right]
\label{EEE}
\eeq
we have that
\beq
{\bbox E}^{(k)}(1,2) = \frac{1}{N} \sum_{ij}
\ov{
\lla {\bbox \phi}_i(1) \left(J^k\right)_{ij} {\bbox \phi}_j(2) \rra
}
\; ,
\eeq
here `$\langle \; \cdot \; \rangle$' denotes mean with the measure
(\ref{1.9}). Correspondingly, the staggered distributions can be
encoded as
\beq
{\bbox g}(\lambda;1,2)
\equiv
g(\lambda;t_1,t_2) +
(\ov \theta_2 - \ov \theta_1) \;
\left[\theta_2 \,
\hat{g}(\lambda;t_1,t_2)
+ \theta_1  \,
\hat{g}(\lambda;t_2,t_1)
\right]
\label{susyg}
\; .
\eeq

We shall use a related set of order parameters
\footnote{ The functions ${\bbox {\cal X}}^{(k)}(1,2)$
are the dynamical analogs of the
functions $X_k$ of Ref. \cite{Mepa}.}
\beq
{\bbox {\cal X}}^{(k)}(1,2)
\equiv
\sum_{r=0}^{k} S_{k,r}\; {\bbox E}^{(r)}(1,2)
\; ,
\eeq
where
$S_{k}(z)=\sum_{r=0}^{k} S_{k,r} \, z^r$
are the Chebyshev
polynomials of the second kind, generated by
\beq
\sum_{k=0}^\infty S_k(z) \;  y^k = \frac{1}{(1- y z + y^2)}
\label{generate}
\; .
\eeq
In components, ${\bbox {\cal X}}^{(k)}(1,2)$ reads
\beq
{\bbox {\cal X}}^{(k)}(1,2) \equiv
{\cal X}^{(k)}(t_1,t_2) +
(\ov \theta_2 - \ov \theta_1) \;
\left[
\theta_2  \,
\hat{{\cal X}}^{(k)}(t_1,t_2)
+ \theta_1 \,
 \hat{{\cal X}}^{(k)}(t_2,t_1)
\right]
\; .
\label{YY0}
\eeq

Following exactly the same steps as in Ref. \cite{Mepa}, one gets
\beq
{\bbox {\cal X}}^{(k)}(1,2)
=
\frac{1}{\pi}
\int_{-2}^2 d\lambda \; \sqrt{1 - \frac{\lambda^2}{4}} \; S_k(\lambda) \;
{\bbox g}(\lambda;1,2)
\; ,
\label{serie}
\eeq
{\it i.e.} each component
${\cal X}^{(k)}$ and  $\hat{{\cal X}}^{(k)}$ is the
coefficient of
the expansion of $g$ and $\hat{g}$ in the polynomials $S_k$.

One can now show \cite{Mepa} that
the ${\bbox {\cal X}}^{(k)}$ are obtained from
\beq
{\bbox {\cal X}}^{(k)}(1,2)
=
\beta^{k} \; {\bbox Q}^{k+1}(1,2)
\; ,
\label{YY}
\eeq
or disentagling the superspace notation
\beqa
{\cal X}^{(k)}(t_1,t_2)
&=&
\frac{1}{\pi}
\int_{-2}^{2} d \lambda \;
\sqrt{1-\frac{\lambda^2}{4}} \;
S_k(\lambda) \; g(\lambda;t_1,t_2)
= \beta^k \, C^{(k+1)}(t_1,t_2)
\\
\hat{{\cal X}}^{(k)}(t_1,t_2)
&=&
\frac{1}{\pi}
\int_{-2}^{2} d \lambda \;
\sqrt{1-\frac{\lambda^2}{4}} \;
S_k(\lambda) \; \hat{g}(\lambda;t_1,t_2)
= \beta^k \, G^{(k+1)}(t_1,t_2)
\; .
\eeqa

We are left with the task of calculating the powers of the superorder
parameter ${\bbox Q}$. Before dealing with this, let us give a compact form
for ${\bbox g}$; using Eq. (\ref{generate}) and the orthogonality
properties of the Chebyshev polynomials we obtain
\beq
{\bbox g}(\lambda;1,2)
=
\left[
{\bbox Q}
\left(
{\bbox \delta} -  \beta \lambda {\bbox Q} + \beta^2 {\bbox Q}^2
\right)^{-1}
\right] (1,2)
\; .
\label{rel}
\eeq
Products and inverses are as in Eq. (\ref{powers})
and the identity is defined as \linebreak
${\bbox \delta}(1-2) \equiv (\theta_2 - \theta_1)
(\ov \theta_2 - \ov \theta_1) \delta(t_2-t_1)$.
The relation (\ref{rel}) is valid for all times.
It is purely a consequence of the
mean-field limit and the (super)symmetries of the problem,
we have not yet used at all the dynamical solution.

In the following sections we shall concentrate on the
long-times regime (\ref{lim1}). We shall express the results
rather than in terms of the times, in terms of the value the auto-correlation
function takes at those times.

\vv
\section{Powers of ${\bbox Q}$}
\renewcommand{\theequation}{\thesection.\arabic{equation}}
\setcounter{equation}{0}
\label{sec3}
\vv

We now calculate the powers ${\bbox Q}^k$ for large times. Until explicitly
noted our calculation is not particular to the SK model but only relies
on the assumptions made in Ref. \cite{Cuku2} for the long times dynamics
of mean-field spin-glasses.

For any three large times the auto-correlations satisfy
`triangle relations':
\beq
C(t_{max},t_{min})
=
f \left(
C(t_{max},t_{int}), C(t_{int},t_{min})
\right)
\; .
\label{trian0}
\eeq
The function $f$ is an associative composition law.

We also have that
\beq
G(t_1,t_2)
=
\frac{\partial F[C(t_1,t_2)]}{\partial t_2}
=
X_d[C(t_1,t_2)] \; \frac{\partial C(t_1,t_2)}{\partial t_2}
\; .
\label{xeq0}
\eeq

Eq. (\ref{xeq0}) define $X_d[C]$ and $F[C]$ (the latter up to a constant).
It says that the violation of the FDT theorem for the non-equilibrium
dynamics of spin-glasses is determined by a function $X_d[C]$ that
depends on the times exclusively through $C(t_1,t_2)$.

This scenario has been proposed to analyse the large-times
dynamics of the mean-field spin-glass
models.
The solution of the dynamical problem for a particular model
gives explicit expressions for $X_d$, $F$ and $f$ \cite{Cuku1,Cuku2}.

\vv

In the Appendices we shall show that the structure
(\ref{trian0})-(\ref{xeq0}) carries through to ${\bbox Q}^k$.
The reasoning is general and does not depend on the model.
The main steps are the following:
We first show that $C^{(k)}$ depends on the times only through
$C(t_1,t_2) $:
\beq
C^{(k)}(t_1,t_2)=C^{(k)}[C(t_1,t_2)]
\label{formula1}
\eeq
The triangle relation for $C^{(k)}$ can be read from
\beq
C(C^{(k)}(t_{max},t_{min}))
=
f \left(
C(C^{(k)}(t_{max},t_{int})), C(C^{(k)}(t_{int},t_{min}))
\right)
\; ,
\label{trian1}
\eeq
{\it i.e.} the new triangle relation is isomorphic to the old one.

Relation (\ref{xeq0}) then maps into
\beq
G^{(k)}(t_1,t_2)
=
\frac{\partial F^{(k)}[C^{(k)}(t_1,t_2)]}{\partial t_2}
=
X_d^{(k)}[C^{(k)}(t_1,t_2)] \; \frac{\partial C^{(k)}(t_1,t_2)}{\partial t_2}
\label{xeq1}
\eeq
In the Appendices we also show for the SK model that
$X_d^{(k)}$ is obtained through
\beq
 X_d^{(k)} (C^{(k)}[C])= X_d[C]
\; .
\label{yy}
\eeq
\vv

Of particular importance are the values of the
correlations $C=a^*_i$ that are `fixed points' of $f$
\beq
f(a^*_i,a^*_i)=a^*_i
\; .
\eeq
Equation (\ref{trian1}) implies that fixed points corresponding to $C$
are mapped into fixed points corresponding to $C^{(k)}$.

The fixed  points separate the range of auto-correlations in `discrete scales'
\cite{Cuku2}.
Under very general (model independent) assumptions,
the relation $f$ between two fixed points is ultrametrical
\beq
f(a^*_i,a^*_j)= \mbox{min} (a^*_i,a^*_j)
\eeq
but not so the relation between values of the auto-correlation
that are not fixed points and belong to the same discrete scale.

In general,
it turns out that the function $C^{(k)}(C)$, when evaluated in the
fixed points $a^*_i$ is related to the ultrametric ansatz in replica
space as follows: Let $Q_{ab}$  be an ultrametric matrix
with elements $q_r$ associated with blocks of sizes $X_r$.
We compute the matrix power $[Q^k]_{ab}$, and consider its elements
(say, $q^{(k)}_r$)
associated with blocks of size $X_r$.
Then, the functional $q^{(k)}[q]$ coincides with the dynamical
functional $C^{(k)}[C]$.

This relationship (at this point purely kinematical)
between powers of static and dynamical
order parameters holds only for `fixed point' values of $C$. The values
of $q$  that are not contained as entries of the ultrametric matrix
correspond to values of $C$ intermediate between fixed points,
{\it i.e.} within discrete scales
\footnote{
 Let us note, in passing, that the correspondence we have just described
is an example of a more general connection between static replica and
dynamic SUSY treatments. Indeed this connection holds not only for powers
of the order parameters, but for a wide class of functionals $H[{\bbox Q}]$
\cite{Cufrkume}.};
for these auto-correlation values there is no replica counterpart within the
ultrametric ansatz.

\vv
\subsection{SK Model}
\vv

For the SK problem in zero magnetic field
\footnote{
This solution has been obtained for $T$ slightly below
the critical temperature $T_c$. We expect it to hold for all temperatures
below $T_c$.},
the solution of the mean-field dynamical equations
yields a dense set of fixed points of $f(C,C)$
in the interval $[0,q_{EA}]$, plus
an isolated fixed point $C(t,t)=1$. The value $q_{EA}$ is the
Edwards-Anderson parameter, and the interval $(q_{EA},1]$ corresponds
to the `FDT' (discrete) scale. For times associated with auto-correlations
in this interval $X_d(C)=1$ and FDT holds.
Instead, for large times associated
to $C$ in $[0,q_{EA}]$, FDT is modified as in Eq. (\ref{xeq0})
by a non-trivial factor $X_d[C]$. The function $X_d[C]$
is part of the solution to the mean-field equations of motion.
\vspace{1cm}

To obtain the explicit form of the powers ${\bbox Q}^k$
it is useful to separate the FDT discrete scale writing ${\bbox Q}(1,2)$ as
\beq
{\bbox Q}(1,2)
=
{\bbox Q}_{FDT}(1,2) + {\bbox {\cal Q}}(1,2)
\label{sep}
\; .
\eeq

The FDT term ${\bbox Q}_{FDT}(1,2)$
has entries that satisfy
\beqa
C_{FDT}(t_1,t_2)
&=&
C_{FDT}(t_1-t_2)
\; ,
\label{FDTc}
\\
G_{FDT}(t_1,t_2)
&=&
\frac{\partial C_{FDT}(t_1-t_2)}{\partial t_2}
\label{FDTg}
\; .
\eeqa
The function $C_{FDT}(\tau)$, $\tau \equiv t_1-t_2$, is a rapidly
(with respect to the variation of ${\bbox {\cal Q}}$)
decreasing function; $C_{FDT}(0) = 1-q_{EA}$ and
$C_{FDT}(\infty) = 0$.
It is the output of the Sompolinsky-Zippelius dynamics
`within a valley' \cite{Sozi}.
Operator powers of ${\bbox Q}_{FDT}$
have entries that verify Eqs. (\ref{FDTc}) and (\ref{FDTg})
and are relevant in the same time-region.

The ${\bbox {\cal Q}}$ function varies slowly; $\cl(t_1,t_1) = q_{EA}$
and $\cl(t_1,t_f) = 0$ if $t_1 >> t_f$.

In the operator product
${\bbox Q}_{FDT} \; {\bbox {\cal Q}}$, the operator
${\bbox Q}_{FDT}$ acts as the identity
${\bbox \delta}(1-2)$
times $1-q_{EA}$ \cite{Ku}.

The separation (\ref{sep}) is the dynamic counterpart of the separation
of the - static - replica matrix $Q_{ab}$,
\beq
Q_{ab} = (1-q_{EA}) \delta_{ab} + {\cal Q}_{ab}
\eeq
where ${\cal Q}_{ab}$ has $q_{EA}$ in the diagonal.

\vv

In order to compute ${\bbox {\cal X}}^{(k)}$ we use that,
for long times,
\beqa
{\bbox Q}^{k}(1,3)
&=&
\sum_{l=0}^k
\;
\left(
\begin{array}{c}
k
\nn
l
\end{array}
\right)
\;
\int d2 \;
{\bbox Q}_{FDT}^{k-l}(1,2) \; {\bbox {\cal Q}}^l(2,3)
\nn
&=&
{\bbox Q}_{FDT}^k(1,3) + \left( (1-q_{EA}) {\bbox \delta}
+ {\bbox {\cal Q}} \right)^k(1,3) - ((1-q_{EA}) {\bbox \delta} )^k (1,3)
\; .
\label{eqQ}
\eeqa
This relation allows to write $\left\{ C^{(k)}, G^{(k)} \right\}$,
the entries of ${\bbox Q}^k$, in terms of
$\left\{ C^{(l)}_{FDT}, G^{(l)}_{FDT} \right\}$ and
$\left\{ {\cal C}^{(l)}, {\cal G}^{(l)} \right\}$,
the entries of ${\bbox Q}^l_{FDT}$ and ${\bbox {\cal Q}}^l$, respectively.

In Appendix A we give expressions for the entries $C^{(k)}_{FDT}$ and
$G^{(k)}_{FDT}$ of ${\bbox Q}^k_{FDT}$.
The explicit form of $C^{(k)}_{FDT}(t_1,t_2)$
has no analog in the ultrametric ansatz for the replica approach,
except for the
values at equal times and at times such that $C=q_{EA}$ ({\it i.e.} at
the limits of the FDT `discrete' scale).
$C^{(k)}_{FDT}$ is also a rapidly decreasing function which falls from
$(1 - q_{EA})^k$ at equal times to zero at widely separated times.

In Appendix B we calculate the entries $\cl^{(k)}$ and $\gl^{(k)}$
of ${\bbox {\cal Q}}^k$ for large times.

\vv

Using these results we
are in a position to express $C^{(k)}$ and $G^{(k)}$ for all ranges of
times:
\begin{itemize}
\item
For large and widely separated times $t_1, t_2$
such that $C(t_1,t_2) < q_{EA}$, we compute the sum in Eq. (\ref{eqQ})
to get, in terms of $C(t_1,t_2)$,
\beqa
C^{(k)} [C]
&=&
\frac{
\left( 1 - q_{EA} - F[0] \right)^k -
\left( 1 - q_{EA} - F[C] \right)^k
}
{ X[\cl] }
\nn
& & -
\int_{X[0]}^{X[C]}
\frac{dx'}{{x'}^2} \;
\left[
\left( 1 - q_{EA} - F[C(x')] \right)^k -
\left( 1 - q_{EA} - F[0] \right)^k
\right]
\;
\label{cc}
\eeqa
and
\beq
G^{(k)}(t_1,t_2)
=
X_d[C] \;
\frac{
\partial C^{(k)}(C)}{\partial t_2}
\; .
\label{gg}
\eeq

\item
For large and close times such that $C > q_{EA}$
\beqa
C^{(k)}(t_1,t_2)
&=&
C^{(k)}_{FDT}(t_1-t_2)    +
C^{(k)}(q_{EA}^-)
\; ,
\\
G^{(k)}(t_1,t_2)
&=&
\frac{\partial C^{(k)}(t_1,t_2)}{\partial t_2}
\; ,
\label{e11}
\eeqa
with $C^{(k)}(q_{EA}^-)$  from Eq. (\ref{cc}).

In particular we shall need the result for equal times
\beq
C^{(k)}(t_1,t_1)
=
(1-q_{EA})^k
+
\cl^{(k)}(q^-_{EA})
\; .
\label{eqq}
\eeq
\end{itemize}

\vv
\section{Expressions for the staggered auto-correlation}
\renewcommand{\theequation}{\thesection.\arabic{equation}}
\setcounter{equation}{0}
\label{sec4}
\vv

Expressions (\ref{e11}) and (\ref{eqq}), together with  (\ref{cc}) and
(\ref{gg})
are all that is needed to calculate the staggered auto-correlation
and response functions at long times.
In order to make contact with the results of Ref. \cite{Mepa}
we make a change of variables:
\beqa
\Delta[X_d]
&\equiv&
F[0] - F[C(X_d)] \; \theta(X_M - X_d)
\\
I
&\equiv&
F[0] + q_{EA}
\eeqa
where $X_M= X_d[q_{EA}^-]$.

Inverting equation (\ref{serie}), using (\ref{generate}) and the
low-temperature phase result $\beta(1-I)=1$, after some algebra we obtain
the staggered auto-correlation at long equal times
\beqa
g(\lambda)
&\equiv&
\lim_{t \rightarrow \infty} \lim_{N \rightarrow \infty}
g(\lambda,t,t)
\nn
&=&
\frac{1}{\beta (2-\lambda)} \;
\left[ \; 1+
  \int_0^1
\frac{dX_d}{X_d^2} \,
\left(
\frac{(\beta \Delta(X_d))^2}
{1-\lambda (1+\beta \Delta(X_d))+(1+\beta \Delta(X_d))^2}
\right)
\;
\right]
\; .
\label{brf}
\eeqa
The staggered auto-correlation $g(\lambda,C)$, between
two large and widely separated times $t_1,t_2$ chosen such that
$C(t_1,t_2)=\cl<q_{EA}$ is given by
\beqa
g(\lambda,\cl)
&\equiv&
\;\;
\lim_{t_1 \rightarrow \infty, C(t_1,t_2)=\cl }
\;\;
\lim_{N \rightarrow \infty}
g(\lambda,t_1,t_2)
\nn
&=&
g(\lambda)-\frac{1}{\beta (2-\lambda)} \;
\left[1
+ \frac{(1-\beta (1-q_{EA}))^2}{1-\lambda \beta (1-q_{EA})+
\beta^2 (1-q_{EA})^2}
\right.
\nn
& &
\;\;\;\;\;\;\;\;\;\;\;
+
\left.\int_{X_d(\cl)}^{X_d(q_{EA}^-)}
\frac{dX_d'}{X_d'^2}
\left(
\frac{(\beta \Delta(X_d'))^2}
{1-\lambda (1+\beta \Delta(X_d'))+(1+\beta \Delta(X_d'))^2}
\right)
\right]
\label{brf1}
\; .
\eeqa
Both these last expressions are valid for the low-temperature phase.

We now note that {\em for the SK model}
the functions $X_d(C)$ for the dynamics and the usual function $X(q)$
of the replica treatment coincide at all temperatures.
Furthermore, the diagonal values $Q^k_{aa}$ and $C^{(k)}(t_1,t_1)$
also coincide. This also implies the equality of the functions $\Delta$
and $I$.

Hence, we have just  proven that for the long and equal times the
dynamic staggered spin auto-correlation (\ref{brf})
coincides with the static one obtained in Ref. \cite{Mepa}.

Furthermore, the staggered auto-correlation $g(\lambda,\cl)$ coincides with
the static one
\footnote{
One can extend the equilibrium  calculation of Ref. \cite{Mepa}
to this case by considering the
entry of the replica matrices
${(1 - \beta \lambda Q + \beta^2 Q^2 )^{-1}}_{ab}$
corresponding to a pair of replicas having mutual overlap $Q_{ab}=\cl$.}
computed with configurations belonging to two equilibrium states
with mutual overlap $\cl$.

\vv

Finally, let us show that, both statically and dynamically, $g(\lambda)$
contains all the information needed to reconstruct $P_{()}(q)$.
To this end we define
\beq
t(X)
\equiv
\frac{1+(1+\beta \Delta(X))^2}{1+\beta \Delta(X)}
\eeq
and
\beq
h(\lambda)
\equiv
\beta \, (2-\lambda) \, g(\lambda) -1
\; .
\eeq
The functions $t(X)$ and $q(X)$ both have a plateau for the same values of
$X \in (X_M,1)$.

Equation (\ref{brf}) becomes
\beq
h(\lambda)= \int_0^{X_M} \frac {dX}{X^2} \; \frac{t(X)-2}{t(X)-\lambda}
+ \left( \frac{1}{X_M} -1 \right) \; \frac{t(X_M)-2}{t(X_M)-\lambda}
\; .
\eeq
Having excluded the plateau in $t(X)$, we can change variables in the integral
to obtain
\beq
h(\lambda)= \int_2^{t_M} \frac{\mu(t') \; dt'}{t'-\lambda}
+ \left( \frac{1}{X_M} -1 \right) \; \frac{t_M-2}{t_M-\lambda}
\eeq
where $t_M \equiv t(X_M)$ and $\mu(t)=(t-2)X^{-2}(t)\;dX/dt$.

This is an electrostatic problem with positive charges.
The  determination of $\mu(t),X_M,t_M$ can in principle be done in a unique
way: the analytical continuation of $h(\lambda)$ from the
interval $-2<\lambda<2$  yields the `charge density'
$\mu(t)$, the magnitude of the `discrete charge' and its position $t_M$.

The knowledge of $\mu(t),X_M,t_M$ then allows to calculate $X(t)$ as
\beq
\frac{1}{X(t)}= \int_t^{t_M} \frac{\mu(t')}{t'-2} dt' + \frac{1}{X_M(t_M)}
\; .
\eeq
Hence, we have shown that $g(\lambda)$ contains all the information needed
to obtain $X(q)$.

\vv
\section{Numerical Simulations}
\renewcommand{\theequation}{\thesection.\arabic{equation}}
\setcounter{equation}{0}
\label{sec5}
\vv

We have performed Montecarlo simulation of the dynamics of
a system with $N=996$ spins at temperature $T=0.3$.

We have calculated the distribution of overlaps for two copies of the system
relaxing from different initial configurations,
at times $t=600,2000,10000$ Montecarlo sweeps.
At these times the system is well out of equilibrium as shown by
the form of the overlap distribution. This eventually takes the
form of the static $P(q)$ (except for finite-size corrections)
at  equilibrium, but is only bell-shaped at the times considered
(see Fig. 1).


\bc
{\bf Figure 1.}
Overlap distributions for $t=600,2000,10000$ Montecarlo sweeps.
The dashed line shows the analytical equilibrium $P(q)$.
\ec

\vspace{1cm}

Figure 2 shows the
equal-times staggered auto-correlation times the density of
eigenvalues, $\rho(\lambda)g(\lambda,t,t)$,
at times $t=600,2000,10000$, together with the analytical result for
the equilibrium $\rho(\lambda)g(\lambda)$.
We notice that the convergence to a curve that coincides with
the equilibrium one is very fast,
even in a situation manifestly  out of equilibrium ({\it cfr.} Fig.1).

In particular, the time dependent energy density is given by
\beq
e(t)
=
\int d\lambda \, \lambda \, \rho(\lambda) \, g(\lambda; t,t)
\; .
\eeq
Hence, the equivalence $\lim_{t \rightarrow \infty} g(\lambda;t,t)
= g(\lambda)$ ensures the equivalence of the asymptotic energy and
the equilibrium one, a result that we have also checked numerically.


\bc
{\bf Figure 2.}
The equal-times staggered auto-correlation distribution $g(\lambda;t,t)$
times the density of eigenvalues $\rho(\lambda)$, for $t=600$ (diamonds),
$t=2000$ (crosses) and $t=10000$ (squares)
Montecarlo sweeps.
The continuous line shows the analytical result for the statics and
for long but finite times.
\ec

\vv
\section{Discussion}
\renewcommand{\theequation}{\thesection.\arabic{equation}}
\setcounter{equation}{0}
\label{sec6}
\vv

The partition function of the SK model is dominated by the
low-lying states. The out of equilibrium dynamics never reaches any of
these states:
there is never a situation of `effective' dynamical
equilibrium in which the system is trapped forever in one of these states
ignoring the rest of the phase-space  and satisfying FDT and time-translational
invariance.

Indeed, as time passes, the  top
evolution of the system
slows down more and more but it is
never completely trapped. In Ref. \cite{Cuku2} it was pointed out that
the equality $P_d(q)=P(q)$ implied that an infinite SK system has an
energy density
which goes asymptotically to the equilibrium energy density.
Furthermore, the `width' of the region in which the system
has a fast relaxation
at long times coincides with the `size' $q_{EA}$ of the equilibrium states.

This already points to a similarity between  the long-time landscape and the
(different) region that dominates the partition function.
The results in this paper suggest that this similarity is much deeper:
Consider  the relaxation at two large times $(t_1,t_2)$.
Because of weak ergodicity breaking \cite{Bo},
given $t_1$ we can always choose
$t_2>t_1$ such that the auto-correlation $C(t_2,t_1)$ between the
configurations
 $\sigma_i(t_1)$ and $\sigma_i(t_2)$
at those times is any given value $\cl$.
  If we now compute the staggered auto-correlation  distribution
$g(\lambda,t_1,t_2)$ for those configurations we obtain the same distribution
we would have obtained with configurations chosen from two equilibrium
states at mutual distance $\cl$.

This result is quite surprising, since we know that the system
 {\em is not in any equilibrium state} at
 times $t_1$ or  $t_2$, however long;
it will
eventually leave the neighbourhood of $\sigma_i(t_1)$ and
$\sigma_i(t_2)$ never to return.

If we now keep the configuration at times $t_2$ and let the system
evolve up to a time $t_3$ such that again $C(t_2,t_3)=\cl$
 we obtain
the same form for the staggered autocorrelation $g(\lambda,t_2,t_3)$.
Note however, that because the system slows down,
$t_2-t_1<t_3-t_2$ if $\cl<q_{EA}$.

The picture that this seems to suggest is that the geometry
of phase space seen at different long-times is similar in every respect,
except that the relevant barriers found at larger times are higher, thus
slowing down the system.

We expect that this similarity between the equilibrium and the
long-time out of equilibrium regions of phase-space will hold for models
that do not have a `threshold' level below which the system cannot go.
More precisely, we expect this similarity to hold for models with a
continuous set of correlation scales; {\it e.g.} the SK model and the
model studied in Refs. \cite{Frme,Frme2}, but we do not expect it hold
in the $p$-spin spherical model of Ref. \cite{Cuku1}.

Finally, let us remark that
the good agreement between the numerical calculation
of $g(\lambda,t,t)$ for large $t$ and the static $g(\lambda)$
constitues a rather detailed test of the solution of the out of equilibrium
dynamics for this model \cite{Fe}.

\vspace{2cm}
\appendix{{\large{Appendix A}}}
\setcounter{equation}{0}
\renewcommand{\theequation}{A.\arabic{equation}}
\vspace{1cm}

In this appendix we give an expression for ${\bbox Q}^k_{FDT}$.
Let us first note that the power of a FDT - supersymmetric -
operator is itself FDT \cite{Ku}.
{}From Eq. (\ref{ceq}) we then have
\beqa
C^{(k)}_{FDT}(t_1-t_3)
&=&
\left[
C^{(k-1)}_{FDT}(t_1-t_2) \,  C_{FDT}(t_3-t_2)
\right]^{t_3}_{0} \nn
&+& \int_{t_3}^{t_1} dt_2 \;
C_{FDT}(t_2-t_3) \, \frac{\partial C_{FDT}^{(k-1)}
(t_1-t_2)}{\partial t_2}
\; .
\label{sab}
\eeqa
Beacause the time-difference $t_1-t_3$ is in a one to one relation
with $C_{FDT}(t_1-t_3)$, Eq. (\ref{sab}) proves Eq. (\ref{formula1})
for the FDT regime.

The value of $G_{FDT}^{(k)}(t_1-t_3)$ is obtained as
\beq
G_{FDT}^{(k)}(t_1-t_3)=
 \frac{\partial C_{FDT}^{(k)}
(t_1-t_3)}{\partial t_3}
\; .
\eeq
This says that $X^{k)}_d = 1$ for $C$ in the FDT regime, and hence
is of the form (\ref{xeq1}).

In this paper we only need the value of
$C^{(k)}_{FDT}(t_1,t_1)=C^{(k)}_{FDT}(0) $. This is easily obtained by putting
$t_3=t_1$ in Eq. (\ref{sab}).
In this way we obtain
\beq
C^{(k)}_{FDT}(t_1,t_1)
=[C_{FDT}(t_1,t_1)]^k
=(1 - q_{EA})^k
\; .
\label{diag}
\eeq

{}From (\ref{sab}) it is easy to see that ${\bbox Q}^{(k)}_{FDT}$ is
also small for very different times.

\vspace{2cm}

\appendix{{\large{Appendix B}}}
\setcounter{equation}{0}
\renewcommand{\theequation}{B.\arabic{equation}}
\vspace{1cm}

In this Appendix we analyse the properties of ${\bbox Q}^k$ for long and
widely separated times, for which
${\bbox Q} = {\bbox {\cal Q}}$ though ${\bbox Q}^k \neq {\bbox {\cal Q}}^k$.
We first study ${\bbox {\cal Q}}^k$ and then the properties for ${\bbox Q}^k$
will follow from linearity (see Eq. (\ref{eqQ})).

We analyse $\cl^{(k)}$, $X_d^{(k)}$ and $F^{(k)}$.
We first demonstrate by induction that $\cl^{(k+1)}$
depends exclusively on $\cl$
\beq
\cl^{(k+1)}(t_1,t_2) = \cl^{(k+1)}(\cl(t_1,t_2))
\; .
\eeq
Then we show also by induction that
the relation between $\gl^{(k+1)}$
and $\cl^{(k+1)}$ maintains the form (\ref{xeq1});
there exist $F^{(k+1)}$ and $X_d^{(k+1)}$ that verify
\beq
\gl^{(k+1)}(t_1,t_2)
=
\frac{\partial F^{(k+1)}[\cl(t_1,t_2)]}{\partial t_2}
=
X_d^{(k+1)}[\cl(t_1,t_2)] \; \frac{\partial \cl^{(k+1)}(t_1,t_2)}{\partial t_2}
\; .
\label{xkeq}
\eeq
Finally we explicitly compute $\cl^{(k)}$, $X_d^{(k)}$ and $F^{(k)}$
in terms of $\cl$ for the SK model.
The $F^{(k)}$ are defined up to a constant, we fix it by imposing
$F^{(k)}({\cal C}(t_1,t_1))=0$ ({\it e.g. } for $k=1$, $F(q_{EA})=0$).

Let us define the (rather badly behaved) `inverse' of $f$
\beq
\cl(t_{max},t_{min})
=
f \left(
\cl(t_{max},t_{int}), \cl(t_{int},t_{min})
\right) \Longrightarrow
\cl(t_{int},t_{min})
=
\ov f \left(
\cl(t_{max},t_{int}), \cl(t_{max},t_{min})
\right)
\label{trian}
\eeq
The function $\ov f$ is written in such a way that
its second argument is
always smaller than the first one.

We start by assuming
\beqa
\cl^{(k)}(t_1,t_2)
\; ,
&=&
\cl^{(k)}(\cl(t_1,t_2))
\\
\gl^{(k)}(t_1,t_2)
&=&
\frac{\partial F^{(k)}[\cl^{(k)}(t_1,t_2)]}{\partial t_2}
=
X_d^{(k)}[\cl^{(k)}(t_1,t_2)]
\; \frac{\partial \cl^{(k)}(t_1,t_2)}{\partial t_2}
\; ,
\eeqa
with $F^{(k)}[\cl^{(k)}(t_1,t_1)] = 0$.
Eq. (\ref{ceq}) then reads
\beqa
\cl^{(k+1)}(t_1,t_3)
&=&
-
\,
\cl^{(k)} \left[ \cl(t_1,0) \right]
\,
F \left[ \cl(t_3,0) \right]
-
\int_{\cl(t_1,0)}^\cl
dy \;
\frac{\partial \cl^{(k)}(y)}{\partial y}
\;
F\left[ \ov f(\cl,y) \right]
\nn
& &
+
\int_{\cl(t_1,0)}^\cl dy \;
X_d^{(k)}\left[ \cl^{(k)}(y) \right]
\;
\frac{\partial \cl^{(k)}(y)}{\partial y}
\;
\ov f(\cl,y)
\nn
& &
+
\int_\cl^{q_{EA}}
dy \;
X_d^{(k)} \left[\cl^{(k)}(y) \right]
\;
\frac{\partial \cl^{(k)}(y)}{\partial y}
\;
\ov f(y,\cl)
\label{ceq2}
\eeqa
where $\cl \equiv \cl(t_1,t_3)$.

It is easy to see that, for all $\ov f$ such that
$\ov f(x,y) \propto y$, $C(t_1,0)= 0 \Rightarrow C^{(k)}(t_1,0) = 0$,
$\forall k$.
In these cases ${\cal C}^{(k+1)}$ reads
\beqa
\cl^{(k+1)}(\cl)
&=&
-
\int_0^\cl
dy \;
\frac{\partial \cl^{(k)}(y)}{\partial y}
\;
F\left[ \ov f (\cl,y) \right]
+
\int_0^\cl dy \;
X_d^{(k)}\left[ \cl^{(k)}(y) \right]
\;
\frac{\partial \cl^{(k)}(y)}{\partial y}
\;
\ov f(\cl,y)
\nn
& &
+
\int_\cl^{q_{EA}}
dy \;
X_d^{(k)} \left[\cl^{(k)}(y) \right]
\;
\frac{\partial \cl^{(k)}(y)}{\partial y}
\;
\ov f(y,\cl)
\; .
\label{ceq3}
\eeqa
This is
a time-reparametrization invariant equality and
${\cal C}^{(k)}$ depends on $t_1$ and $t_3$ only
through  $\cl$.
\vspace{1cm}

We now demonstrate that Eq. (\ref{xkeq}) holds $\forall k$.
Eq. (\ref{geq}) implies
\beqa
G^{(k+1)}(t_1,t_3)
&=&
\frac{\partial }{\partial t_3}
\int_{t_3}^{t_1}
dt_2  \;
\frac{ \partial F^{(k)} \left[ \cl^{(k)}(t_1,t_2) \right]}{\partial t_2}
\;
F\left[\ov f \left( \cl(t_1,t_2), \cl(t_1,t_3) \right) \right]
\label{geq2}
\eeqa
Then, we can identify
\beqa
F^{(k+1)}(t_1,t_3)
&=&
\int_{t_3}^{t_1}
dt_2  \;
\frac{ \partial F^{(k)} \left[ \cl^{(k)}(t_1,t_2) \right]}{\partial t_2}
\;
F\left[\ov f \left( \cl(t_1,t_2), \cl(t_1,t_3) \right) \right]
\label{geq3}
\eeqa
choosing the integration constant to be zero.
Now, using $\cl^{(k)}(t_1,t_2)=\cl^{(k)}(\cl(t_1,t_2))$
\beq
F^{(k+1)}(t_1,t_3)
=
\int_\cl^{q_{EA}}
dy  \;
\frac{ \partial F^{(k)} \left[ \cl^{(k)}\right]}{\partial \cl^{(k)}}
\;
\frac{ \partial \cl^{(k)}(y)}{\partial y}
\;
F\left[\ov f \left( y, \cl \right) \right]
\label{geq4}
\; .
\eeq
The rhs only depends on $\cl$ and
then $F^{(k+1)}(t_1,t_3) = F^{(k+1)}[\cl^{(k+1)}]$ and
$X_d^{(k)}(t_1,t_3) = X_d^{(k+1)}[\cl^{(k+1)}]$.

\vspace{1cm}

The derivation up to this point is general, it only depends on the assumptions
 of Ref. \cite{Cuku2}.
For the SK model $\cl(t_1,0) = 0$ for long enough time $t_1$ and
the ultrametric dynamical
relation between auto-correlation functions
\beq
\ov f (x,y) = \min(x,y) = y
\label{ult}
\;
\eeq
holds \cite{Cuku2}.
Then $C^{(k)}(t_1,0) = 0$, $\forall k$ and
\beqa
\cl^{(k+1)}(\cl)
&=&
-
\int_0^\cl
dy \;
\left(
\frac{\partial \cl^{(k)}(y)}{\partial y}
\;
F\left[ y \right]
+
F^{(k)}\left[ \cl^{(k)}(y) \right]
\right)
\; .
\label{ceq4}
\eeqa

We can also solve Eq. (\ref{geq4})
using the ultrametric relation (\ref{ult}). We obtain
\beqa
F^{(k)} \left[ \cl^{(k)}(\cl) \right]
&=&
-
\left( - F\left[ \cl \right] \right)^k
\label{Feq}
\eeqa

We now solve Eq. (\ref{ceq4}) for $\cl^{(k+1)}$.
Its derivative w.r.t. $\cl$ is
\beq
\frac{ \partial \cl^{(k+1)}(\cl)}{\partial \cl}
=
-
\frac{ \partial \cl^{(k)}(\cl)}{\partial \cl}
\;
F[\cl]
-
F^{(k-1)}[\cl^{(k-1)}]
\; .
\label{parc}
\eeq
Using Eq.(\ref{Feq}) we get the recursive equation
\beq
w^{(k+1)} = w^{(k)} + 1
\eeq
with
$w^{(k)} \equiv (-F[\cl])^{k-1} \; \partial \cl^{(k)}(\cl)/\partial \cl$.
The solution is
\beqa
\frac{\partial \cl^{(k)}(\cl)}{\partial \cl}
&=&
k \; (-F[\cl])^{k-1}
\label{cres1}
\\
\cl^{(k)}(\cl)
&=&
- \int_0^\cl dy \; \frac{\partial (-F[y])^k}{\partial y} \frac{1}{X_d[y]}
\label{cres2}
\eeqa

We now  obtain
$X_d^{(k)}[\cl^{(k)}(\cl)]$ in terms of $X_d[\cl]$. Derivating Eq. (\ref{Feq})
w.r.t. $\cl$
\beq
X_d^{(k)}[\cl^{(k)}(\cl)] \;
\frac{\partial \cl^{(k)}(\cl)}{\partial \cl}
=
k
\;
\left(-F[\cl]\right)^{k-1}
\;
X_d[\cl]
\;
\eeq
and inserting the result in Eq. (\ref{cres1}) we obtain
\beq
X_d^{(k)}[\cl^{(k)}(\cl)]
=
X_d[\cl]
\label{xxxxx}
\; .
\eeq

We have obtained these results ${\cal C}^{(k)} = {\cal C}^{(k)} ({\cal C})$
and Eq. (\ref{xxxxx}) for the powers ${\bbox {\cal Q}}^k$.
Since these hold for every power $k$ and $C^{(k)}$ is a linear
combination of these powers (see Eq. (\ref{eqQ}))
we have proven Eqs. (\ref{formula1}) and (\ref{yy}).

\end{document}